\documentclass[conference,10pt]{IEEEtran}
\IEEEoverridecommandlockouts
\usepackage{authblk}

\usepackage[utf8]{inputenc}
\usepackage{braket}
\usepackage{tikz}
\usepackage{circuitikz}
\ctikzset{logic ports=ieee}
\usepackage{latexsym}
\usepackage{qcircuit}
\usepackage{threeparttable}
\usepackage{amsfonts}
\usepackage{multirow}
\usepackage{algorithm}
\usepackage{algorithmic}
\usepackage{url}

\usepackage{amsmath,amssymb,amsthm}
\usepackage{subfigure}
\usepackage{flowchart}

\usepackage[normalem]{ulem}
\newcommand {\qvar} {qv}
\def \Rm#1{\mbox{\rm #1}}
\def \lsem      {\raise1pt\hbox{\Rm {[\kern-.12em[}}}
\def \rsem      {\raise1pt\hbox{\Rm {]\kern-.12em]}}}
\def \sem#1{\mbox{\lsem$#1$\rsem}}

\newcommand {\iif} {\mathbf{if}}
\newcommand {\fii} {\mathbf{fi}}

\usepackage{xcolor}
\newtheorem{defn}{Definition}

\newtheorem{exam}{Example}

\title{Equivalence Checking of Dynamic Quantum Circuits}

\author[1]{Xin~Hong}
\author[1*]{Yuan~Feng\thanks{*Email:\ \{yuan.feng,sanjiang.li,mingsheng.ying\}@uts.edu.au.}}
\author[1*]{Sanjiang~Li}
\author[1,2,3*]{Mingsheng~Ying}

\affil[1]{Centre for Quantum Software and Information, University of Technology Sydney, Australia}
\affil[2]{State Key Laboratory of Computer Science, Institute of Software, Chinese Academy of Sciences, China}
\affil[3]{Department of Computer Science and Technology, Tsinghua University, China}

\begin{document}
\maketitle

\begin{abstract}
Despite the rapid development of quantum computing these years, state-of-the-art quantum devices still contain only a very limited number of qubits. One possible way to execute more realistic algorithms in near-term quantum devices is to employ dynamic quantum circuits, in which measurements can happen during the circuit and their outcomes are used to control other parts of the circuit. This technique can help to significantly reduce the resources required to achieve a given accuracy of a quantum algorithm. However, since this type of quantum circuits are more flexible, their verification is much more challenging. In this paper, we give a formal definition of dynamic quantum circuits and then propose to characterise their functionality in terms of ensembles of linear operators. Based on this novel semantics, two dynamic quantum circuits are equivalent if they have the same functionality. We further propose and implement two decision diagram-based algorithms for checking the equivalence of dynamic quantum circuits. Experiments show that embedding classical logic into conventional quantum circuits does not incur significant time and space burden. 
%our algorithm is as effective as the existing algorithms for equivalence checking of conventional quantum circuits.

% In this way, the number of qubits required could be significantly reduced.
%
%to measure between operations and split the whole functionality into smaller ones that are connected by classical control based on the measurement results. %to decrease the number of qubits needed. %

%Despite the rapid development of quantum computing these years, quantum hardware is still limited both in scale and accuracy. One possible way to execute more realistic  algorithms in the near future quantum devices is to use dynamic quantum circuit techniques; that is, to split the circuit into many pieces and connect them using classical control to decrease the qubits needed and obtain other benefits. However, since this type of quantum circuit is more flexible, their verification is much more challenging. In this paper, we give a formal definition of dynamic quantum circuits and propose an algorithm for checking the equivalence of two dynamic quantum circuits. Experiments show that our algorithm is as effective as the existing  algorithms for checking two conventional quantum circuits.
\end{abstract}

\begin{IEEEkeywords}
Quantum circuits, quantum measurements,  dynamic quantum circuits, equivalence checking
\end{IEEEkeywords}

\section{Introduction}
%\lsj{I will revise this part Thursday morning.}
%A possible outline of the introduction:

%\begin{enumerate}
%    \item Deferred measurement principle is ... However, sometimes it is more desired to use this principle reversely ... \xh{The deferred measurement principle maybe not necessary and I have deleted it from the text.}
    % \item What are dynamic quantum circuits? What are their potential benefits? Short qubit coherence, qubit reuse, shallow circuit depth, less noise, more precise computational results
    % \item Why equivalence checking is important for dynamic quantum circuits? Introduce existing methods for standard quantum circuits and explain why they do not work for dQCs. 
    % \item Contribution of this paper: Formalise the definition of equivalence checking dynamic quantum circuits, provide a short introduction and explain why we introduce two kinds of equivalence checking; 
    % propose a decision diagram-based method for checking the equivalence of dQCs, explain how we extend the existing tdd-based equivalence checking approach. 
%\end{enumerate}

The past five years have witnessed significant breakthroughs in building small-scale quantum devices and the largest quantum computers now have above 50 qubits. It is widely believed that near-term quantum devices will remain very limited in terms of the number of qubits they may have. Furthermore, these quantum devices also suffer from noises and short coherence time. This makes it highly difficult to implement large practical quantum circuits. 

To fully exploit the power of quantum computing in these noisy intermediate-scale quantum (NISQ) \cite{PreskillNISQ} devices, several prominent approaches have been proposed to overcome the tight scale restriction. These include the quantum network approach \cite{Wehner18internet}, which connects multiple small-scale quantum computers, hybrid quantum/classical algorithms (see  \cite{Endo+JPSJ21} for a survey), which join quantum computers with classical computers, and dynamical quantum circuits \cite{ryan2017hardware,corcoles2021exploiting}, which try to build classical logic directly into quantum circuits.

Conventionally, a quantum circuit starts with a set of qubit initialisation, then an ordered sequence of qubit gates, and ends with or without measurements. As pointed out in \cite{corcoles2021exploiting}, in these conventional quantum circuits, classical logic is not performed within the coherence time of the qubits.  

Dynamic quantum circuits are circuits in which measurements can be performed in the middle of the circuit and the measurement results
are then used to control some 
quantum gates to be applied afterwards; early examples include  quantum teleportation \cite{barrett2004deterministic,riebe2004deterministic,steffen2013deterministic,chou2018deterministic} and quantum error correction (QEC) \cite{reinhold2020error,minev2019catch,gottesman2010introduction,paler2017fault}. For QEC, the syndrome of an encoded quantum state is detected through a set of measurements, and a series of quantum gates are then applied to the error quantum state according to the measurement results, and  thus the quantum state is recovered. 

%Different kinds of encoding can be proposed based on various situations, but they all benefits from the dynamic quantum circuit technique.

The dynamic quantum circuit  technique has also been used in extending the lifetime of quantum bits \cite{ofek2016extending}, resetting qubits \cite{riste2012feedback}, realising non-elementary quantum gates \cite{ryan2017hardware,bravyi2019simulation},  generating entanglement  \cite{riste2013deterministic,saira2014entanglement}, and embedding the classical real-time logic into  quantum systems \cite{corcoles2021exploiting,andersen2019entanglement}. Experiments in \cite{corcoles2021exploiting} show that dynamic circuits can indeed offer a `substantial and tangible advantage'  on current noisy quantum hardware.

As dynamic quantum circuits will play a more important and central role in the near-future quantum computing, their verification becomes an imperative problem. Still lacking a formal definition of dynamic quantum circuits, it is not surprise that this problem is completely untouched. This paper provides a first such attempt and focuses on the equivalence checking of dynamic quantum circuits. Similar to the case of classical circuits, equivalence checking is essential to maintain the correctness of a quantum circuit. Moreover, as quantum circuits become larger and larger, they are more and more error-prone. It is necessary to provide automatic tools for checking the equivalence of different quantum circuit designs. For conventional quantum circuits, the decision diagram-based approach plays a prominent role, see, e.g., \cite{niemann2015qmdds,viamontes2007checking,burgholzer2020improved,hong2020tensor}. For two quantum circuits, their equivalence is simply reduced to checking if the corresponding decision diagram representations are identical. However, due to the existence of classical control, it is not clear if all these decision diagram approaches can be directly applied to checking the equivalence of two dynamic quantum circuits. 

In this paper, we first give a formal definition of dynamic quantum circuits and lay a rigorous foundation for our discussion by characterising their functionalities 
%of dynamic quantum circuits 
in terms of ensembles of linear operators. Then we formally define the notion of equivalence of dynamic quantum circuits and present two special cases, called m- and q-equivalences, which cover most, if not all, existing realistic dynamic quantum circuits. The m-equivalence focuses on the measurement results while the q-equivalence cares only about the output quantum states. The quantum phase estimation (QPE) algorithm and QEC are, respectively, representative examples of the application scenarios of the two equivalence definitions. Then, we propose and implement two algorithms for checking the m- and q-equivalences, based on the tensor decision diagram (TDD) \cite{hong2020tensor}. As a data structure, TDD provides a compact and canonical representation for tensors and, in particular, quantum circuits. It can be used in many design automation tasks, e.g., simulation \cite{hong2020tensor} and equivalence checking \cite{hong2021approximate}, for quantum circuits. In order to represent dynamic quantum circuits, what we need to do is to (a) represent measurements and classically controlled gates as tensors and construct their TDDs and (b) represent Boolean functions as TDDs. This is natural as TDDs are generalisations of binary decision diagrams \cite{bryant1992symbolic} and tensors are, in a sense, generalisations of Boolean functions.  

%TDD can be easily extended to represent Boolean functions and, hence, classical combinational circuits.

%To make it more suitable for the dynamic quantum circuits, we extend the tensor representation for classical circuits, and then the the whole dynamic quantum circuits can be represented by a TDD.

Our main contributions are summarised as follows:
\begin{enumerate}
    \item A formal  definition of dynamic quantum circuits and the characterisation of their functionalities in terms of ensembles of linear operators;
    \item A formal definition of dynamic quantum circuit equivalence and two special cases, viz., m- and q-equivalences,  covering most real-world application scenarios;
    \item Implementation of two TDD-based algorithms for checking the m- and q-equivalences.
    %of dynamic quantum circuits.
%    \item The TDD representation of the classical circuits are discussed and used to check the equivalence of dynamic quantum circuits.
\end{enumerate}

In the remainder of this paper, we first recall some basic concepts of quantum computing and tensor networks   in Sec.~\ref{sec:background}, and then present our formal definitions of dynamic quantum circuits and their equivalence  in Sec.~\ref{sec:dqc}. Algorithms for checking the m- and q-equivalences of dynamic quantum circuits are described in Sec.~\ref{sec:equivalence checking}, followed by experiments and numerical results in Sec.~\ref{sec:experiments} and conclusion in  Sec.~\ref{sec:conclusion}.

\section{Backgrounds}\label{sec:background}

In this section, we recall some basic concepts from quantum computing and tensor networks. For quantum computing, we adopt notations from \cite{NC00, ying2016foundations}. Interested readers may consult  \cite{markov2008simulating, biamonte2019lectures} for more details in tensor networks.

\subsection{Quantum Circuits}
% In the classical computation, digital circuits are made from logic gates acting on Boolean variables, while in the quantum computation, 
A (conventional) quantum circuit is a series of quantum gates, which are modelled by unitary matrices, acting on qubit (quantum bit) variables. For each qubit $q$, we write $\mathcal{H}_{q}$ for its state Hilbert space, which is two-dimensional. Using the Dirac notation, a pure state of $q$ is represented by $|\psi\rangle=\alpha_0|0\rangle+\alpha_1|1\rangle$ with complex numbers $\alpha_0$ and $\alpha_1$ satisfying $|\alpha_0|^2+|\alpha_1|^2=1$.  A sequence $\overline{q}=q_1,...,q_n$ of distinct qubit variables is called a quantum register. Its state Hilbert space is the tensor product $\mathcal{H}_{\overline{q}}=\bigotimes_{i=1}^n\mathcal{H}_{q_i}$, which is $2^n$-dimensional. Thus, an $n$-qubit state can be also represented by an $2^n$-dimensional vector $|\psi\rangle=(\alpha_0,...,\alpha_{2^n-1})^T$. A unitary transformation on $\overline{q}$ is modelled by a $2^n\times 2^n$ unitary matrix $U$. This transformation is also called an $n$-qubit quantum gate and denoted as $G\equiv U[\overline{q}]$. We use the notation $\qvar(G)$ to represent the qubits that the gate is operated on. A quantum circuit can be formed by a sequence of quantum gates: $C\equiv G_1;\ldots;G_d$, and the quantum register of $C$ is denoted
$\qvar(C)=\bigcup_{i=1}^{d}\qvar(G_i)$. 

To read out the information obtained by running a quantum circuit, measurement are sometimes applied at the end of a circuit. Let $\overline{q}=q_1,...,q_n$ be a quantum register. A measurement on $\mathcal{H}_{\overline{q}}$ is a collection $\{ M_m \} \subseteq \mathcal{L}(\mathcal{H}_{\overline{q}})$ of operators
satisfying the normalisation condition: $\sum_m{M_m^{\dagger}M_m}=I_{\mathcal{H}_{\overline{q}}}$, where $I_{\mathcal{H}_{\overline{q}}}$ is the identity matrix in $\mathcal{H}_{\overline{q}}$ and $M_m$ (the index $m$ stands for the measurement outcome) are called measurement operators. If the state of a quantum system is $\ket{\psi}$ immediately before the measurement, then, for each outcome $m$, the probability that $m$ occurs is $p(m) = \bra{\psi}M_m^{\dagger}M_m\ket{\psi}$ and the state of the system after observing $m$ is $\ket{\psi_m} = M_m\ket{\psi}/\sqrt{p(m)}$.

The basic measurement used in this paper is the measurement in the computational basis, which is formed by two operators $M_0=\ket{0}\bra{0}$, $M_1=\ket{1}\bra{1}$ for a single qubit. If the qubit before the measurement was in state $\ket{\psi}=\alpha_0 \ket{0} +\alpha_1 \ket{1}$, then the probability of obtaining outcome $m\in\{0,1\}$ is $p(m) = \bra{\psi}M_m^{\dagger}M_m\ket{\psi} = |\alpha_m|^2$, and the state after measurement is $M_m\ket{\psi}/\sqrt{p(m)}=\ket{m}$.

\begin{figure}
\centerline{
\Qcircuit @C=1em @R=0.9em {
\lstick{\ket{0}}  & \gate{H} &\qw      &\ctrl{2}  &\gate{H} &\ctrl{1} &\qw &\meter \\
\lstick{\ket{0}}  & \gate{H} &\ctrl{1} &\qw       &\qw      &\gate{S^{\dagger}}&\gate{H}&\meter\\
\lstick{\ket{v}} & {/}\qw      &\gate{U} &\gate{U^2}&\qw      &\qw&\qw&\qw\\
}
}
\caption{Quantum circuit for 2-qubit Phase Estimation. The wires from top to bottom represent qubits $q_1$, $q_2$, and $r$ respectively.}
\label{exp-for-quantum-circuit}
\end{figure}
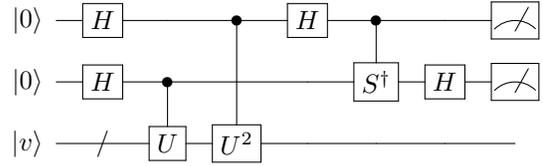

\begin{exam}\label{ex-q-circuit}
Depicted in Fig.~\ref{exp-for-quantum-circuit} is a quantum circuit which implements the 2-qubit Phase Estimation~\cite{NC00}, where 
$$
H=\frac{1}{\sqrt{2}}{
\left[ \begin{array}{cc}
1 & 1\\
1 &-1\\
\end{array} 
\right ]},\quad
S^{\dagger}={
\left[ \begin{array}{ccc}
1 & 0\\
0& -i \\
\end{array}
\right ]}
,$$
 $U$ is a unitary matrix, and $\ket{v}$ is an eigenstate of $U$; that is, $U\ket{v} = e^{2\pi i \phi}\ket{v}$ for some $\phi\in [0,1)$. Here the two measurements at the end are both in the computational basis, and the observed outcomes give the best two-bit approximation of $\phi$ with high probability. In particular, if $\phi = 0.\phi_1\phi_2$ is the binary representation of $\phi$, then we will obtain $\phi_1$ in $q_1$ and $\phi_2$ in $q_2$ with probability 1.
 %If a gate $G$ on some qubit $q_1$  is connected by a black dot on  another qubit $q_2$, this denotes that $G$ is controlled by qubit $q_2$; that is, $G$ will be applied iff qubit $q_2$ is in state $\ket{1}$.
\end{exam}

\subsection{Tensor Networks}
%In addition to the basic concepts of quantum circuits, we also need to introduce some basic concepts of tensor networks. 

A \emph{tensor} is a multidimensional linear map associated with a set of indices. %For each assignment of the values of these indices, the tensor maps them to a number. 
In this paper, we assume that each index takes value in $\{0,1\}$ and tensors take values from $\mathbb{C}$, the field of complex numbers. That is, a tensor $\phi$ with index set $I=\{x_1, \ldots, x_n\}$ is a mapping from $\{0,1\}^I$ to $\mathbb{C}$. It is also denoted by $\phi_{x_1\ldots x_n}$ or $\phi_{\vec{x}}$. The value of $\phi$ under an assignment $\{x_i \mapsto  a_i, 1\leq i\leq n\}$ is denoted by $\phi_{x_1\ldots x_n}(a_1, \ldots, a_n)$, or $\phi_{\vec{x}}(\vec{a})$, or even $\phi(\vec{a})$ for simplification. The number $n$ of the indices of a tensor is called its \emph{rank}. Scalars, 2-dimensional vectors, and $2\times 2$ matrices are tensors with rank 0, 1, and 2, respectively. 

The most important tensor operation is \emph{contraction}. For any two tensors, their contraction is a tensor obtained by summing up over shared indices. For example, let $\gamma_{x_1, x_2}$ and $\xi_{x_2, x_3}$ be two tensors which share a common index $x_2$. Then their contraction is a new tensor $\phi_{x_1, x_3}$ with $\phi_{x_1, x_3}(a_1, a_3)=\sum_{a_2 \in \{0,1\}}{\gamma_{x_1, x_2}(a_1, a_2)\cdot \xi_{x_2, x_3}}(a_2, a_3)$.
% \begin{equation}\label{eq:contdef}
% 	\phi_{x, y}(a, b)=\sum_{c \in \{0,1\}}{\gamma_{x, z}(a, c)\cdot \xi_{y, z}}(b, c).
% \end{equation}
When both $\gamma$ and $\xi$ are $2\times 2$ matrices,  contraction is exactly matrix multiplication.

Another useful tensor operation is \emph{slicing}, which corresponds to the cofactor operation of Boolean functions. Let $\phi$ be a rank $n+1$ tensor. Then its slicing w.r.t. $x=c$ for index $x$ and $c\in\{0,1\}$ is a rank $n$ tensor. For example, let $\phi$ be a tensor with index set $I=\{x, x_1,\ldots,x_n\}$. The slicing of $\phi$ w.r.t. $x = c$ is a tensor $\phi|_{x=c}$ over $I'=\{x_1, \ldots, x_n\}$ given by $\phi|_{x=c}(\vec{a}):= \phi(c, \vec{a})$,
% \begin{align}
% \phi|_{x=c}(\vec{a}):= \phi(c, \vec{a})
% \end{align}
for any $ \vec{a} \in \{0,1\}^n$. We call $\phi|_{x=0}$ and $\phi|_{x=1}$ the \emph{negative} and \emph{positive} slicing of $\phi$ with respect to $x$, respectively. 

Quantum states and quantum gates can both be represented by tensors. For example, the quantum state $\ket{0}$ can be represented by a rank 1 tensor $\phi_{x}$ with $\phi_{x}(0)=1, \phi_{x}(1)=0$, and the $H$ gate in Example \ref{ex-q-circuit} is represented by the rank 2 tensor with  $\phi_{x_1,x_2}(00)=\phi_{x_1,x_2}(01)=\phi_{x_1,x_2}(10)=1/\sqrt{2},\phi_{x_1,x_2}(11)=-1/\sqrt{2}$. Clearly, each quantum circuit can be represented by the tensor obtained by contracting all its gates and input/output states.

%In addition, a tensor is called a COPY tensor if it equals 1 when all its indices are assigned the same value and equals 0 otherwise.

A set of tensors can be connected to form a network. A \emph{tensor network} is an undirected graph $G=(V, E)$ with zero or multiple open edges, where each vertex $v$ in $V$ represents a tensor and each edge a common index associated with the two adjacent tensors.
% \begin{defn}[Tensor networks]\label{def-tensor-net}
% A tensor network is an undirected graph $G=(V, E)$ with zero or multiple open edges, where each vertex $v$ in $V$ represents a tensor and each edge a common index associated with the two adjacent tensors.
% \end{defn}
Tensors sharing the same index should be contracted, and this contraction leads to a rank $m$ tensor if there are $m$ open edges in the tensor network. In this way, quantum circuits are special tensor networks with every node representing a quantum gate or a quantum state.
% and the edges representing the connections \yf{contraction?} of quantum gates or quantum states. 
% When there is no input state assigned in the tensor network, contraction of this tensor network will give a rank $2n$ tensor which represents the unitary matrix corresponding to the quantum circuit. Here, $n$ is the number of qubits of the circuit. If all the input states have been included in the tensor, contracting the tensor network will give a rank $n$ tensor which represents the output state of this circuit operating on these input states.

%Henceforth, the notions of tensor and quantum circuit will be used interchangeably if there is no ambiguity. 

\subsection{Tensor Decision Diagrams}
The most commonly used equivalence checking methods for conventional quantum circuits are based on decision diagram~\cite{niemann2015qmdds,viamontes2007checking,burgholzer2020improved,hong2020tensor}. %One of the newest proposed decision diagram is called the tensor decision diagram (TDD) \cite{hong2020tensor}.
The current paper %follows the approach of
adopts the tensor decision diagrams (TDDs) proposed in \cite{hong2020tensor} which can represent tensors and,  thus, quantum circuits, in a canonical and compact way. Given a tensor $\phi$, the root node $r_\phi$ of its TDD $\Phi$ is labelled with an index of $\phi$ and the two successors of $r_\phi$ represent the negative (0-successor) and positive (1-successor) slicing of $\phi$ w.r.t. this index. The value of $\phi$ under an assignment is obtained by multiplying the weights of the corresponding path in $\Phi$.

%Given a tensor $\phi$, its TDD $\Phi$ is constructed recursively by first introducing a root node $r_\phi$, which is labelled with an index of $\phi$, and then attaching to $r_\phi$ as two successors the two TDDs that represent the negative (0-successor) and positive (1-successor) slicing of $\phi$ respect to this index 

%In the TDD representation, a tensor is represented by a node labelled with an index of this tensor and two successors representing the negative (0-successor) and positive (1-successor) slicing of this tensor respect to this index. The value of the tensor can be obtained by multiplying the weights of a corresponding path in the tensor decision diagram.

%two successors  representing . The value of the tensor is  obtained by multiplying the weights of a corresponding path in the tensor decision diagram.

\begin{figure}
    \centering
    \includegraphics[width=0.33\textwidth]{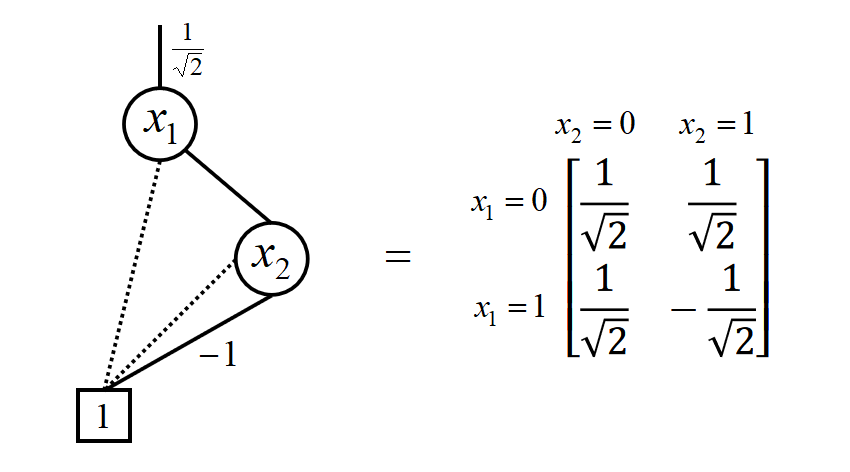}
    \caption{TDD of the $H$ gate.}
    \label{fig:TDD_exp}
\end{figure}

\begin{exam}
Fig. \ref{fig:TDD_exp} shows the TDD of the $H$ gate. In this diagram, dotted lines correspond to the 0-successors, and solid lines the 1-successors. The weight $1/\sqrt{2}$ on the incoming edge of the root  is called the weight of the TDD. The dotted line of the root node (labelled with index $x_1$) leads to the terminal node 1, meaning that when the index $x_1$ takes value 0, the tensor will take value $1/\sqrt{2}\times 1=1/\sqrt{2}$. All 1-weights are omitted in the diagram.
\end{exam}

% \begin{figure}
%     \centering
%     \subfigure[]{
%     \begin{circuitikz}
%     \draw[->] (0,0.8)--(0,0.4);
%     \draw (0,0) circle (0.4);
%     \node at (0, 0.2) [anchor=north]{$x_1$};
%     \draw (1,-1) circle (0.4);
%     \node at (1, -0.8) [anchor=north]{$x_2$};
%     \draw (-0.3,-2) rectangle (0.4,-2.7);
%     \node at (0.05, -2.1) [anchor=north]{$1$};
%     \draw[->] (0.2,-0.38)--(0.65,-0.8);
%     \draw[dashed,->] (0,-0.4)--(-0.1,-2);
%     \draw[dashed,->] (0.6,-1.2)--(-0.05,-2);
%     \draw[->] (1,-1.4)--(-0.02,-2);
%     \end{circuitikz}
%     }
%     \subfigure[]{
    
%     }
%     \subfigure[]{
    
%     }
%     \caption{Examples of TDD and BDD, (a) an TDD representation of a rank 2 tensor, while only $\phi_{x_1,x_2}(11)=1$, (b) the BDD of the logical AND operation, (c) the TDD representation of the logical AND operation.}
%     \label{fig:TDD_exp}
% \end{figure}

A TDD $tdd$ is determined by its weight $tdd.weight$ and the root node $tdd.root$, every node in a TDD has an index $node.index$ and, except the terminal node, two successors $node.succ_0$ and $node.succ_1$. Two TDDs $tdd_1$ and $tdd_2$ are \emph{identical}  iff $tdd_1.weight=tdd_2.weight$ and $tdd_1.root=tdd_2.root$.
%, and they are \emph{identical up to a global phase} iff $tdd_1.node=tdd_2.node$ and $|tdd_1.weight|=|tdd_2.weight|$. 
As TDD provides canonical representation for conventional quantum circuits~\cite{hong2020tensor}, to check if two such circuits are equivalent, we first calculate their  TDDs and then compare their weights and nodes. If their TDDs are identical, then these two quantum circuits are equivalent.

\section{Dynamic Quantum Circuits}\label{sec:dqc}

Informally speaking, a dynamic quantum circuit is a conventional circuit with measurements in the middle and the subsequent circuits can depend on the measurement outcomes \cite{corcoles2021exploiting}.
To establish a rigorous foundation for verification of such circuits,  we next give a formal definition,
%, in which future gates 
% \yf{gates?}\xh{They use states in the original paper, but I think that maybe circuits are better.}
%depend on outcomes of measurements that happen beforehan
generalising the notion of quantum circuits with measurements in \cite{ying2016foundations} by introducing more flexible control flows.

\subsection{Dynamic Quantum Circuits}

\begin{defn}[Dynamic Quantum Circuits]\label{def-dQC}
A dynamic quantum circuit (dQC) is defined inductively as follows:
\begin{enumerate}
    \item Each conventional quantum circuit $C$ acting on qubits $\overline{q}$ is a dQC, and $\qvar(C)= \overline{q}$;
    \item Let $M$ be the measurement in computational basis of $\mathcal{H}_{\overline{r}}$ and, for each $0\leq i< 2^t$, $C_i$ a dQC with $\overline{r} \cap \qvar(C_i) = \emptyset$. Let $f : \{0,1\}^{|\overline{r}|} \rightarrow \{0,1\}^t$ be a Boolean function. Then $$C\equiv\ \iif\ (\Box i \cdot f(M[\overline{r}]) = i \rightarrow C_i)\ \fii$$  is a dQC, and $\qvar(C)=\bigcup_i \qvar(C_i)$.
    \item If $C_1$ and $C_2$ are dQCs, then so is $C_1;C_2$ and $\qvar(C_1;C_2)= \qvar(C_1)\cup \qvar(C_1)$.
\end{enumerate}
\end{defn}

Intuitively, the construct in Clause 2 of the above definition denotes a dQC which first measures the qubits in $\overline{r}$ according to the measurement $M$. If $f$ maps the measurement outcome to $m$, then $C_m$ is chosen to be executed subsequently. Furthermore, Clause 3 describes sequential composition of dQCs.

Similar to conventional quantum circuits, a dynamic circuit is sometimes accompanied with a fixed input state for some of its qubits. Furthermore, a specific set of qubits will be discarded at the end and their quantum states are not included in the circuit output. To incorporate this situation, in the following we denote a dQC as a tuple $$(C[\overline{q}],\ket{\psi}, \overline{i}, \overline{o})$$
where $\overline{q}=\qvar(C)$, and $\overline{i}$ and $\overline{o}$ are both subsets of $\overline{q}$ indicating the (principal) input qubits and output qubits, respectively, and $\ket{\psi}\in \mathcal{H}_{\overline{q}\backslash \overline{i}}$ is its fixed input state.

\begin{defn}[Functionality of dQCs]\label{def-sdQC}
	The functionality of a dQC $C$, denoted $\sem{C}$, is an ensemble of linear operators on $\mathcal{H}_{\qvar(C)}$ defined inductively as follows:
	\begin{enumerate}
		\item If $C$ is a conventional quantum circuit, then $\sem{C} = \{U\}$ where $U$ is the unitary operator computed by $C$ in the usual way;
		\item Let $$C\equiv\ \iif\ (\Box i \cdot f(M[\overline{r}]) = i \rightarrow C_i)\ \fii$$ 
		and $\sem{C_i} = \{F_{i,k} : k\in J_i\}$ for each $i$.
		Then  $\sem{C}$ is an ensemble
		\[
		\left\{\ket{j}_{\overline{r}}\bra{j}\otimes F_{f(j), k}: 0\leq j< 2^{|\overline{r}|}, k\in J_{f(j)}
		\right\}.
		\] 
		\item If $C_1$ and $C_2$ are dQCs with $\sem{C_1} = \{E_i : i\in I\}$ and $\sem{C_2} = \{F_j : j\in J\}$, then $$\sem{C_1;C_2} = \{F_jE_i : i\in I, j\in J\}.$$
	\end{enumerate}
	
	Finally, for dQCs where certain input qubits are with fixed initial states and certain output qubits are discarded, the functionality is defined as follows. Let $\sem{C} = \{F_j : j\in J\}$. Then $\sem{(C[\overline{q}],\ket{\psi}, \overline{i}, \overline{o})}$ is a super-operator from $\mathcal{H}_{\overline{i}}$ to $\mathcal{H}_{\overline{o}}$ such that for any $\rho\in \mathcal{D}(\mathcal{H}_{\overline{i}})$,
	\[
	\sem{(C[\overline{q}],\ket{\psi}, \overline{i}, \overline{o})}(\rho) = \mathrm{tr}_{\overline{q}\backslash \overline{o}} \left[ \sum_{j\in J}  F_j\left(\ket{\psi}\bra{\psi}\otimes \rho\right)F_j^\dag\right].
	\]
	Here $\mathrm{tr}_{\overline{q}\backslash \overline{o}}$ is the \emph{partial trace} operation which discards the $\overline{q}\backslash \overline{o}$ register and takes the reduced quantum state in the remaining part (the $\overline{o}$ register).
\end{defn}

\begin{figure}
\centerline{
\Qcircuit @C=.7em @R=.4em @! {
 & \qw & \qw & \ctrl{1} &
\gate{H} & \meter & \control \cw\\
\lstick{\ket{0}} & \qw & \targ & \targ & \qw &
\meter & \cwx\\
\lstick{\ket{0}} & \gate{H} & \ctrl{-1} & \qw &
\qw & \gate{X} \cwx & \gate{Z} \cwx &
 \qw   \gategroup{1}{2}{3}{5}{.7em}{--}
}
}
\caption{Quantum circuit for Teleportation. The wires from top to bottom represent qubits $q$, $q_1$, and $q_2$ respectively.}
\label{exp-for-dQC}
\end{figure}
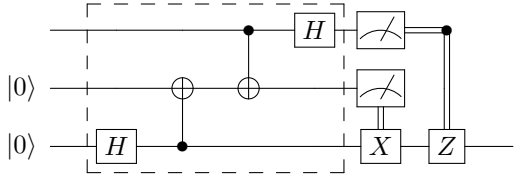

\begin{exam}\label{ex-dQC}
Fig. \ref{exp-for-dQC} gives an example of dynamic quantum circuit. This circuit is the well known circuit for quantum teleportation which transfers a qubit by just sending two classical bits of information. Let the (conventional) circuit in the dashed box be $C_0[q,q_1,q_2]$, and $D_0 \equiv I$, $D_1 \equiv X[q_2]$, $D_2 \equiv Z[q_2]$, and $D_3 \equiv ZX[q_2]$. Then the whole dynamic circuit can be written as $(C[q,q_1,q_2], \ket{00}_{q_1,q_2}, q, q_2)$ where
\[
C\equiv C_0;  \iif\ (\Box i \cdot M[q,q_1] = i \rightarrow D_i)\ \fii
\]
and $M = \sum_{i=0}^3 \ket{i}\bra{i}$ is the 2-qubit measurement %according to
in the computational basis. It is easy to check that for any $\rho\in \mathcal{D}(\mathcal{H}_q)$, $
\sem{(C[q,q_1,q_2], \ket{00}_{q_1,q_2}, q, q_2)}(\rho) = \rho.$

% For this circuit, the quantum gates (two $H$ gates and two $CNOT$ gates) before the measurement form a conventional quantum circuit. Then, two measures are applied on the first and the second qubits of the circuit, and a $X$ and $Z$ gate are applied to the third qubit of the circuit based on the measurement results. These two measurements can also be seen as a single measurement applied on the first two qubits and the gates $I$, $X$, $Z$, $ZX$ will be applied to the third qubit of the circuit according to four different measurement results.
\end{exam}

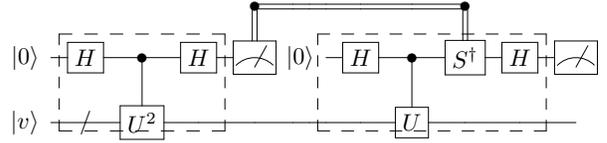
\begin{figure}
\centerline{
\scalebox{0.9}{
\Qcircuit @C=0.7em @R=1.3em {
& & & &\control&\cw&\cw&\cw&\cw&\cw&\control \cw \cwx[1] &&\\
\lstick{\ket{0}}  & \gate{H}&\ctrl{1}   &\gate{H} &\meter \cwx[-1] &&& \lstick{\ket{0}}  & \gate{H}&\ctrl{1}   &\gate{S^{\dagger}}            &\gate{H} &\meter\\
 \lstick{\ket{v}} & {/}\qw  &\gate{U^2} &\qw&\qw&\qw&\qw&\qw&\qw  &\gate{U}   &\qw     &\qw&\qw
  \gategroup{2}{2}{3}{4}{.7em}{--}
  \gategroup{2}{8}{3}{12}{.7em}{--}
  }
  }
  }
\caption{Dynamic quantum circuit for 2-qubit phase estimation. The wires with input $\ket{0}$ from left to right represent qubits $q_1$ and $q_2$, respectively, and the wire on the bottom denotes qubit $r$.}
\label{exp-for-dynamic_PE}
\end{figure}

The circuit in Fig. \ref{exp-for-quantum-circuit} can also be transformed to a dynamic quantum circuit, since the controlled-$S$ gate in the circuit do not change the measurement result of the first qubit. Thus, we can conduct the measurement before the controlled-$S^{\dagger}$ gate and replace it with a classically controlled gate.

\begin{exam}\label{ex-dQC-PE}
Fig.~\ref{exp-for-dynamic_PE} shows the dynamic quantum circuit version of the phase estimation shown in Fig. \ref{exp-for-quantum-circuit}. 
% First, the circuit outside the box (the two $H$ gates and the controlled-$U^2$ gate) is a dQC. Then a measurement is applied on the first qubit. If the measurement result is 0, the dynamic quantum circuit in the box without the $S^{\dagger}$ gate (the two $H$ gates and the controlled-$U$ gate) will be executed, otherwise, the circuit in the box with the $S^{\dagger}$ gate (the two $H$ gates and the controlled-$U$ gate as well as the $S^{\dagger}$ gate) will be executed.
Let the (conventional) circuit in the left dashed box be $C_0[q_1,r]$, the circuit in the right dashed box be $D_1[q_2,r]$, and the circuit in the right dashed box without the $S^\dag$ gate be $D_0[q_2,r]$. Then the whole dynamic circuit can be written as $(C[q_1,q_2,r], \ket{00}_{q_1,q_2}\ket{v}_{r}, \emptyset, q_1q_2)$ where
\[
C\equiv C_0;  \iif\ (\Box i \cdot M[q_1] = i \rightarrow D_i)\ \fii
\]
and $M$ is the 1-qubit computational basis measurement. Note that there is no principal input for this circuit.  It is easy to check that $
\sem{(C[q_1,q_2,r], \ket{00}_{q_1,q_2}\ket{v}_{r}, \emptyset, q_1q_2)} = \ket{\phi_1\phi_2}\bra{\phi_1\phi_2}$ 
%\lsj{$\sem{(C[q,q_1,q_2], \ket{00}_{q_1,q_2}\red{\ket{v}_{r}}, \red{\emptyset}, \red{q_1q_2})} = \ket{x_1x_2}\bra{x_1x_2}$?}
provided that $U\ket{v} = e^{2\pi i0.\phi_1\phi_2}\ket{v}$.
\end{exam}

\subsection{Equivalence of dQCs}

With Definition~\ref{def-sdQC}, we can now define the equivalence of two dQCs as follows.
\begin{defn}[Equivalence of dQCs]
Two dynamic quantum circuits $(C[\overline{q}],\ket{\psi}, \overline{i}, \overline{o})$ and $(C'[\overline{q}'],\ket{\psi'}, \overline{i}, \overline{o})$ with the same principal input and output qubits are equivalent if their functionality are the same; that is, \[
\sem{(C[\overline{q}],\ket{\psi}, \overline{i}, \overline{o})}=  \sem{(C'[\overline{q}'],\ket{\psi'}, \overline{i}, \overline{o})}.
\]
\end{defn}

%\[\sem{(C[\overline{q}],\ket{\psi}, \overline{i}_1, \overline{o}_1)}=  \sem{(C'[\overline{q}'],\ket{\psi'}, \overline{i}_2,\overline{o}_2)},\]
% where there are  bijective mappings $f_i:\overline{i}_1 \to \overline{i}_2$ and $f_o:\overline{o}_1 \to \overline{o}_2$.

In this paper we are particularly concerned with the equivalence of two special classes of dynamic quantum circuits, which cover most real-world applications. In the first class,  all input states are fixed (i.e. $\overline{i} = \emptyset$) and all principal output qubits have been measured during execution of the circuit (the outcomes were or were not used in subsequent circuits). Typical examples include phase estimation in Example~\ref{ex-dQC-PE} and the period finding algorithm. Therefore, the output can be regarded as purely classical information; more precisely, it is merely a probability distribution over $\{0,1,\ldots,2^{|\overline{o}|}-1\}$. We call the equivalence of such dynamic circuits \textbf{m-equivalence}. It is not difficult to see that circuits depicted in Figs.~\ref{exp-for-quantum-circuit} and \ref{exp-for-dynamic_PE} are m-equivalent. 

In the second class of dynamic circuits, measurement outcomes during the execution process were used to control subsequent circuits, and do not constitute a part of the output of the entire circuit. Furthermore,
the output quantum state is independent of the measurement outcomes.
Typical examples in this class include teleportation in Example~\ref{ex-dQC} and error correction. We call the equivalence of such dynamic circuits \textbf{q-equivalence}. 
Again, it is easy to show that the circuit depicted in Fig.~\ref{exp-for-dQC} is q-equivalent to the dynamic circuit $(\mathit{SWAP}[q,q_2], \ket{0}_{q_2}, q, q_2)$ which transfers the state of $q$ to $q_2$ by employing the swap gate.

\section{TDD-based Equivalence Checking}\label{sec:equivalence checking}
In this section, we present our methods and algorithms for checking the m- and q-equivalence of dynamic quantum circuits. The idea is to represent each dynamic quantum circuit as a TDD and then compare if they are identical. To this end, we first consider how to represent measurements and classically controlled gates as tensors, and then consider how to represent Boolean functions as TDD. In this way, we shall have ensured that every element in a dynamic quantum circuit can be represented as a TDD. As a consequence, the TDD representation of the whole dynamic quantum circuit can be obtained by TDD contractions.  

%combined with decision diagrams. Whether we check the m-equivalence or the q-equivalence of two dynamic quantum circuits, the most important thing is to calculate the decision diagram that represent the dynamic quantum circuit.

% \lsj{Reorganise this section. It seems that classical logic is not used in the algorithms. The first subsection concerns regarding measurements and classical controlled gates as tensors; the second then about classical logic; and the third presents the basic algorithms; and the last the optimisation.}

\subsection{Tensor Representation of Measurements and Classically Controlled Gates}
%Decision diagrams have been widely used in the equivalence checking of classical and quantum circuits \cite{viamontes2003improving,niemann2015qmdds,burgholzer2020improved,hong2020tensor}. The decision diagram we use in this paper is the tensor decision diagram (TDD) \cite{hong2020tensor}. To use the tensor decision diagram, we have to ensure that every element in the circuit is transformed to a tensor. As discussed in section \ref{sec:background}, the quantum gates and quantum states can be easily represented by tensors. Thus, what we need to do is to give \textbf{a tensor representation of the measurement and the classically controlled gates}.

Note that measurement in the computational basis of a multi-qubit system can be decomposed into those of the 1-qubit subsystems.
%For convenience, we assume that all measurements are single qubit measurements in the computation basis. 
Furthermore, the probability of measuring a qubit can be read out from the TDD representation of a quantum circuit. Therefore, there is no need to really operate a measurement if it is at the end of the circuit so will not be used to control other parts of the circuit. 

More precisely, we regard each measurement as a rank 2 COPY tensor $\phi_{x,y}=I$ provided that the measurement result is not used to control other parts of the circuit. Otherwise, we regard it as a rank 3 COPY tensor $\phi_{c,x,y}$, where $\phi(000)=\phi(111)=1$ and it equals 0 in all other cases. In this representation, $c$ captures the behaviour of the measurement result and $x,y$ capture the behaviour of the quantum state. When the measurement result is $c=0$, the tensor is exactly $\phi|_{c=0}=\ket{0}\bra{0}=M_0$; and if $c=1$, then it equals to $\phi|_{c=1}=\ket{1}\bra{1}=M_1$. 

A classically controlled-$U$ gate can also be interpreted as a rank 3 tensor $\psi_{c,x^{\prime},y^{\prime}}$, where $\psi|_{c=0}=I$ and $\psi_{c=1}=U$. That is, when the measurement result (i.e., the classical control bit) is 1 the quantum gate $U$ will be applied, and, otherwise, it does nothing. Suppose $M$ is the measurement whose outcomes are used to control $U$. Then the contraction of $\psi$ with the tensor that represents $M$ gives a rank 4 tensor $\ket{0}\bra{0}\otimes I+\ket{1}\bra{1}\otimes U$, which is equivalent to a quantum controlled-$U$ gate.

%We can see that a measurement that has not been used as a controlling part does nothing to the circuit and the 

%The contraction of a measurement the outcomes of which are used as classical control and the classically controlled-$U$ gate gives a rank 4 tensor $\ket{0}\bra{0}\otimes I+\ket{1}\bra{1}\otimes U$, which is equivalent to a quantum controlled-$U$ gate.

The TDD of a dynamic quantum circuit can be obtained by contracting all tensors in the circuit. 
%By comparing the decision diagrams obtained from two circuits, we can determine their equivalence.

% that represent a conventional circuit equivalent to the circuit transformed by the dynamic quantum circuit through the deferred measurement principle. 

\begin{exam}\label{ex-alg}
Consider the circuit in Fig.~\ref{exp-for-dynamic_PE}. The contraction of the first measurement with the classically controlled-$S^\dag$ gate equals to the quantum controlled-$S^\dag$ gate in Fig. \ref{exp-for-quantum-circuit} and the measurement in the end of the circuit is equivalent to an identity matrix. Thus, contracting all these tensors is just equivalent to contracting all tensors in Fig. \ref{exp-for-quantum-circuit} before measurement, which is also equivalent to contracting all tensors in Fig.~\ref{exp-for-quantum-circuit} (including the measurement). This shows that these two dynamic quantum circuits are equivalent.
\end{exam}

\subsection{Representing Boolean Functions as TDDs} \label{subsec:cl_log}
In a dynamic quantum circuit, measurement results are usually sent through a classical combinational circuit and the outputs are used to control the quantum system (see Clause 2 of Definition~\ref{def-dQC} for the formal description).
%processed by a classical computer and then feeding back to the quantum system. 
Thus, we need also provide TDD representations for the classical logic.

% \begin{figure}
%     \centering
%     \begin{circuitikz}[scale=0.7, transform shape]
%     \draw (3,3) node[and port] (and1) {};
%     \node at (1.5, 3.5) [anchor=north]{$x_1$};
%     \node at (1.5, 2.9) [anchor=north]{$x_2$};
%     \node at (4.4, 3.2) [anchor=north]{$y$};
% \end{circuitikz}
%     \caption{The logic AND gate.}
%     \label{fig:log-and}
% \end{figure}

A classical logic gate is in essence  a Boolean function $f: \{0,1\}^n \to \{0,1\}$. 
%Let $x_1,\cdots,x_n$ be the input signals of this logic gate, then $f(x_1,\cdots,x_n)=1$ if the output of this logic gate is 1, and $f(x_1,\cdots,x_n)=0$, when the output of this gate is 0.
%
To represent $f$ as a tensor, we introduce a new index $y$ to describe its output signal. Then, $f$ can be described as a tensor $\phi: \{0,1\}^{n+1} \to \{0,1\}$ with index set $\{x_1,\cdots,x_n,y\}$, where $\phi_{x_1,\cdots,x_n,y}=1$ iff $f(x_1,\cdots,x_n)=y$. We call this the tensor representation of $f$.

For example, consider the logic $AND$ gate whose Boolean function is $f(x_1,x_2)=x_1 \wedge x_2$, which takes value 1 iff both $x_1$ and $x_2$ take values 1. Introducing a new index $y$ to represent the output signal of this gate, $f$ can be represented by a tensor $\phi_{x_1x_2y}$, where $\phi(000)=\phi(010)=\phi(100)=\phi(111)=1$ since $f(00)=f(01)=f(10)=0$ and $f(11)=1$, and the tensor takes value 0 for all other combinations. 

Let $\varphi_{x_1,\cdots,x_n,y_1}$ and $\gamma_{x_1,\cdots,x_n,y_2}$ be the tensor representations of two Boolean functions. Suppose $y_1$ and $y_2$ are used as input signals into another logic gate with tensor representation $\xi_{y_1,y_2,z}$. The contraction of the three tensors on $y_1,y_2$ is
$$
\phi_{x_1,\cdots,x_n,z}=\sum_{y_1,y_2}\varphi_{x_1,\cdots,x_n,y_1} \cdot \gamma_{x_1,\cdots,x_n,y_2} \cdot \xi_{y_1,y_2,z}.
$$
Since, $\varphi,\gamma,\xi$ equal 1 iff $x_1,\cdots,x_n,y_1,y_2,z$ coincide with the input and output behaviour of the Boolean function, thus, the contracted tensor $\phi_{x_1,\cdots,x_n,z}$ equals 1 iff $x_1,\cdots,x_n,z$ coincide with the input-output behaviour of connecting the two outputs $y_1,y_2$ to the inputs of the last logic gate, i.e., $\phi$ is the tensor representation of the combined circuit.

According to this observation, the behaviour of a classical circuit can thus be captured using the contraction of tensor networks. In practice, the Boolean function is often given as a binary decision diagram (BDD) \cite{bryant1992symbolic}. In this case, it can be easily transformed to a TDD by introducing two nodes which are labelled with the output index of the classical circuit and represent the tensors $\phi_y(0)=0, \phi_y(1)=1$ and $\phi_y(1)=0, \phi_y(0)=1$ respectively. 

\begin{figure}
    \centering
    \subfigure[]{
    \includegraphics[width=0.12\textwidth]{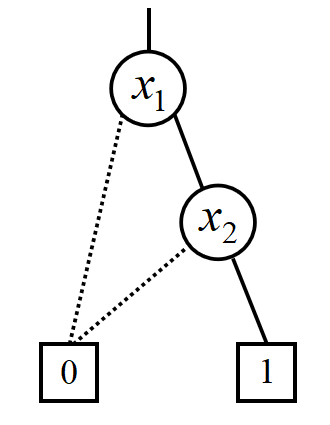}
    }
    \subfigure[]{
    \includegraphics[width=0.11\textwidth]{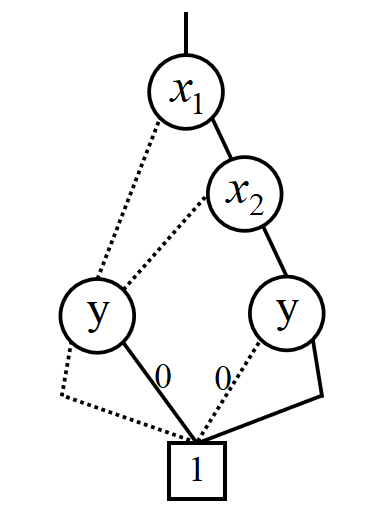}
    }
    \caption{The BDD (a) and TDD (b) representations of the logical AND gate.}
    \label{fig:TDD_extend}
\end{figure}

\begin{exam}
Fig. \ref{fig:TDD_extend} (a) is the BDD representation of the logic $AND$ gate. For this gate, introduce an index $y$ to describe its output, then redirect every edge linked to terminal node 0 to a node labelled with $y$ with 0-successor 1 and 1-successor 0, and redirect every edges linked to terminal node 1 to a node labelled with $y$ with 0-successor 0 and 1-successor 1, such as shown in Fig. \ref{fig:TDD_extend} (b). Then this TDD captures the behaviour of the Boolean function and can be used to do further calculation.
\end{exam}

%Then, for checking the equivalence of dynamic quantum circuits, we need to calculate the 

In this way, the TDD representation of the classical circuit can be calculated from its Boolean function or transformed from its BDD representation. Contracting the TDDs of the classical parts and quantum parts of the dynamic circuit finally gives the TDD of the whole circuit.

\subsection{Basic Algorithms}
The direct method for checking the equivalence of two dynamic quantum circuits is to first calculate their TDD representations and then compare.

%Unlike the exact equivalence, which ask the TDDs to be exactly equal, for the two kinds of equivalence, the TDDs should satisfy certain properties.

{\bf m-Equivalence.} 
For m-equivalence, we have to extract the probabilities of observing different measurement outcomes from the TDD of the dynamic quantum circuit. Suppose the two TDDs to be compared are constructed using the same index order, and all output qubits are on the top of the TDDs. 
%Then, the probability of a certain sequence $s$ of measurement outcomes is exactly the norm of the sub-TDD rooted at the node obtained from the root of the original TDD following the path determined by $s$, with the weight being accumulated along the path. 
For a sequence $s$ of measurement outcomes, let $r_s$ be the node reached from the root of the TDD following the path determined by $s$. The probability of $s$ is exactly the norm of the sub-TDD rooted at $r_s$ weighted by the magnitude of the weight accumulated along the path.
Here the norm of a TDD is obtained by contracting the TDD with its conjugate transpose, which has the same structure but with all corresponding weights conjugated.
% The probability of $s$ is exactly the norm of the sub-TDD rooted at $r_s$ with the weight being accumulated along the path.
% Here the norm of a TDD is obtained by adding up the square of the norms of the weights along all the paths from the root to the terminal. 
Finally, the circuits represented by the two TDDs are m-equivalent if, for each sequence $s$ of measurement outcomes, the corresponding probabilities determined by $s$ in these two TDDs are equal.
% the probability can be computed by iteratively pass the weight of the TDD to its successors until the non-measured output indices are reached. Finally, Here  
Alg.~\ref{alg:is_equivalent} gives the detailed steps of this process. 

%For a certain sequence $s$ of measurement outcomes, let $r_s$ be the node obtained from the root of the original TDD following the path determined by $s$. The probability of $s$ is exactly the norm of the sub-TDD rooted at $r_s$ with the weight being accumulated along the path.

%the probability of a  is exactly the norm of the sub-TDD rooted at the node obtained from the root of the original TDD following the path determined by $s$, with the weight being accumulated along the path.

% Then, this can be done by delete the nodes below the measured index and replace it with a weight on the corresponding edge which equals the inner product of this tensor. The weight then will all be transformed to its norm. Then comparing the normalised TDD, if the two TDD are equal, then this two circuits are m-Equivalent.

\begin{algorithm}
\caption{$m\_eq(tdd_1,tdd_2)$}
\label{alg:is_equivalent}
\begin{algorithmic}
\REQUIRE $tdd_1$, $tdd_2$ are the TDDs of two dynamic circuits
\ENSURE return\ \TRUE\  iff the circuits represented by $tdd_1$ and $tdd_2$ are m-equivalent
\IF{$tdd_1=tdd_2$} 
\RETURN \TRUE
\ENDIF
\STATE $x \leftarrow \min(tdd_1.root.index,\  tdd_2.root.index)$
\IF{$x$ is not a measurement index}
\STATE $N_1,\ N_2 \leftarrow$ norm of $tdd_1$, norm of $tdd_2$
\RETURN $N_1=N_2$
\ENDIF

\FOR{$i=1,2$}
\IF{$tdd_i.root.index=x$}
\STATE $w_i\leftarrow tdd_i.weight$
\STATE $L_i\leftarrow |w_i|\cdot tdd_i.root.succ_0$
\STATE $H_i\leftarrow |w_i|\cdot tdd_i.root.succ_1$
\ELSE
\STATE $L_i, H_i\leftarrow tdd_i$
\ENDIF
\ENDFOR
\RETURN $m\_eq(L_1,L_2) \wedge m\_eq(H_1,H_2)$

% \IF{$N_1=N_2$} 
% \RETURN True
% \ELSE
% \RETURN False
% \ENDIF
\end{algorithmic}
\end{algorithm}

{\bf q-Equivalence.} For q-equivalence, we ignore probabilities of the measurement results and require that the state remained after measurements are independent of the measurement results. We need check if the output state is the same for all input states. If so, the two circuits are q-equivalent. Again, we assume that the measured qubits are on the top of the two TDDs and the two TDDs use the same index order. Alg.~\ref{alg:is_equivalent2} gives the detailed procedure of this process. In the algorithm, we recursively visit the nodes of a TDD and add them to a set if the index of this node is not a measurement index. Since  nodes in the TDD that represent the same tensor share the same address, thus, if the two circuits are q-equivalent, then, finally, the set will become a singleton.
%will be only one node in the set finally 
The algorithm uses this observation to determine the equivalence.

% \begin{algorithm}
% \caption{$q\_eq(tdd_1,tdd_2)$}
% \label{alg:is_equivalent2}
% \begin{algorithmic}
% \REQUIRE $tdd_1$, $tdd_2$ are the TDDs of two dynamic circuits
% \ENSURE return True iff the circuits represented by $tdd_1$ and $tdd_2$ are q-equivalent
% \STATE $nodes \leftarrow empty\_set()$
% \STATE $nodes \leftarrow get\_nodes(tdd_1,nodes)$
% \STATE $nodes \leftarrow get\_nodes(tdd_2,nodes)$
% \RETURN $|nodes|=1$
% \newline
% \STATE \leftline{$//$ get\_nodes(tdd,nodes) subroutine}
% Sub-Alg: $get\_nodes(tdd,nodes)$
% \IF{the smallest index $x$ is an measurement index} 
% \STATE $L\leftarrow x.succ_0$
% \STATE $H\leftarrow x.succ_1$
% \STATE $nodes \leftarrow get\_nodes(L,nodes)$
% \STATE $nodes \leftarrow get\_nodes(H,nodes)$
% \RETURN $nodes$
% \ELSE
% \RETURN $nodes \cup \{tdd.node\}$
% \ENDIF
% \end{algorithmic}
% \end{algorithm}

\begin{algorithm}
\caption{$q\_eq(tdd_1,tdd_2)$}
\label{alg:is_equivalent2}
\begin{algorithmic}
\REQUIRE $tdd_1$, $tdd_2$ are the TDDs of two dynamic circuits
\ENSURE return \TRUE\ iff the circuits represented by $tdd_1$ and $tdd_2$ are q-equivalent
\STATE $nodes \leftarrow get\_nodes(tdd_1)\cup get\_nodes(tdd_2)$
\RETURN $|nodes|=1$
\newline\\

\STATE $/*$ $get\_nodes$ subroutine $*/$

Subroutine $get\_nodes(tdd)$
\STATE $r\leftarrow tdd.root$
\IF{$r.index$ is a measurement index} 
\RETURN $get\_nodes(r.succ_0) \cup get\_nodes(r.succ_1)$
\ELSE
\RETURN $\{r\}$
\ENDIF
\end{algorithmic}
\end{algorithm}

% \begin{algorithm}
% \caption{$q\_eq(tdd_1,tdd_2)$}
% \label{alg:is_equivalent2}
% \begin{algorithmic}
% \REQUIRE $tdd_1$, $tdd_2$ are the TDDs of two circuits
% \ENSURE return True iff the circuits represented by $tdd_1$ and $tdd_2$ are q-equivalent
% \STATE $res1,state1\leftarrow if\_only\_one\_state(tdd_1)$
% \STATE $res2,state2\leftarrow if\_only\_one\_state(tdd_2)$
% \IF{$res1 = False$ or $res2 = False$} 
% \RETURN False
% \ELSE 
% \RETURN $state1=state2$
% \ENDIF
% \newline
% \STATE \leftline{$//$ if\_only\_one\_state(tdd) subroutine}
% Sub-Alg: $if\_only\_one\_state(tdd)$
% \IF{the smallest index $x$ is an measurement index} 
% \STATE $L\leftarrow tdd.node.succ_0$
% \STATE $H\leftarrow tdd.node.succ_1$
% \STATE $res1,state1 \leftarrow if\_only\_one\_state(L)$
% \STATE $res2,state2 \leftarrow if\_only\_one\_state(H)$
% \IF{$res1 = False$ or $res2 = False$} 
% \RETURN False, None
% \ELSE
% \RETURN $state1=state2, state1$
% \ENDIF
% \ELSE
% \RETURN True, tdd
% \ENDIF
% \end{algorithmic}
% \end{algorithm}

\subsection{Optimisation} \label{subsec:opt}

%Since in dynamic quantum circuits, a circuit is usually executed part by part, we can also check the equivalence of the two circuits by partitioning them into many sub-circuits. When we find two equivalent sub-circuits in the two circuits, we can discard them and check the remaining part of the two circuits.

The equivalence of dynamic quantum circuits can be checked in a divide-and-conquer manner. Suppose the circuits are partitioned into corresponding sub-circuits. If two equivalent corresponding sub-circuits are found, we delete them from the circuits and check the remaining parts. The circuit partition scheme has been used in tensor network-based quantum circuit simulation \cite{pednault2017breaking} and TDD calculation \cite{hong2020tensor}. In this paper, 
%based on the form of the two specific kinds of equivalence, 
we consider a simple circuit partition scheme in which the circuit is partitioned qubit-by-qubit. More precisely, we first calculate the two tensors corresponding to the first qubit, compare them and then consider the second qubit and so on. 

For m-equivalence, if the tensors corresponding to a same qubit of the two circuits are identical, they can be discarded. This is because, if such two tensors are identical, then the distributions of measuring this qubit are identical, and discarding them will not influence the equivalence of measuring other qubits.
%
%if the tensors corresponding to a qubit of the two circuits are equal, they can be discarded. This is because, if the tensors corresponding to a qubit are equal, then the distributions of measuring this qubit are equal, and discarding these two tensors will not influence the equivalence of measuring other qubits.
% if the tensors corresponding to every qubit are equal, the distributions of measuring the two qubits are equal.
For q-equivalence, we should ensure that the tensors corresponding to every pair of qubits satisfy the requirement for q-equivalence, that is, the remaining tensor do not depend on the measurement indices, and the tensors of the two circuits are also identical. Then, such pairs of tensors can be discarded.

This qubit-by-qubit partition can help us reduce the number of tensors needed to be considered at the same time. If there are still tensors that cannot be discarded, then they should be collected and contracted and checked using the basic algorithms. %If they obey the corresponding algorithms, then these two circuits are m- or q-equivalent. 
The effectiveness of this optimisation scheme has been confirmed in experiments (see Sec.~\ref{sec:experiments}).

\section{Numerical Results}\label{sec:experiments}
In this section, we evaluate the effectiveness of our algorithms, with the aim to confirm that no significant time and space consumption will be incurred from embedding classical logic into conventional quantum circuits. More precisely, we compare our equivalence checking algorithms for dynamic quantum circuits with the TDD-based equivalence checking method for conventional quantum circuits. In addition, we also implemented the optimised algorithms based on qubit-by-qubit partition and compared them with the basic algorithms. 

{\bf Benchmarks.}
We choose the dynamic quantum circuits of some commonly used algorithms such as quantum Fourier transform (QFT) \cite{griffiths1996semiclassical}, phase estimation (PE) \cite{corcoles2021exploiting} and circuits for error correction \cite{devitt2013quantum}, teleportation \cite{NC00} and state injection \cite{ryan2017hardware}. For each circuit, we compare them with the corresponding conventional quantum circuits given in \cite{NC00}. 
%As there is no conventional circuit provided for teleportation, we generate one using the principle of deferred measurement \cite{NC00}; that is, to move the measurement to the end of the circuit and replace the classically controlled gate with a quantum controlled gate. 
We then check m-equivalence for the quantum Fourier transform and phase estimation circuits and check q-equivalence for the error correction, teleportation, and state injection circuits.

% For QFT, we choose the circuits given in \cite{NC00} as conventional quantum circuits, while the corresponding dynamic quantum circuits are given in \cite{griffiths1996semiclassical}. For PE, we randomly choose a single qubit diagonal matrix as the target unitary, and the corresponding conventional and dynamic quantum circuits are given in \cite{NC00} and \cite{corcoles2021exploiting}, respectively. 

% On the other hand, to randomly generate dynamic quantum circuits, we choose the gate and measurement randomly from $\{X,Y,Z,H,S,T,CX,Measurement\}$ and connecting them to dynamic quantum circuits and corresponding conventional quantum circuits. For the $QFT$ and $PE$, we only consider the m-equivalence, and for the q-equivalence, we only care about q-equivalence, for the random circuits, we consider both the types of equivalence.

{\bf Implementation.} In our experiments, for each pair of conventional quantum circuit $C$ and its corresponding dynamic circuit $C'$, we calculate the TDD representations of  $C$ and $C'$ and check their equivalence by comparing these two TDDs. To this end,
%for each quantum algorithm we calculate the TDD representations of the conventional quantum circuit and its corresponding dynamic circuit and check their equivalence by comparing these two TDDs. To this end,
%In order to perform tensor network contraction in checking the equivalence of tensors representing the two circuits in question, 
we employ the TDD package provided in~\cite{hong2020tensor}.
% which are suitable for contracting tensor networks and provided a canonical representation for tensors.
All the verification algorithms are implemented using Python3 and the experiments are conducted on a laptop with Intel i7-1065G7 CPU and 8GB RAM.    
% \footnote{Source codes available at \url{https://anonymous.4open.science/r/EC-for-Dynamic-Quantum-Circuits-B8FE/}}
\subsection{Comparison with EC of Conventional Quantum Circuits}
{\bf Baseline.} We compare the time and memory consumption of our verification algorithms with the TDD-based equivalence checking method proposed in~\cite{hong2020tensor} for conventional quantum circuits.
%, 
%that is, by first calculating the decision diagrams of the two conventional quantum circuits and then checking if they are identical, 
%which is the same method we used for checking the equivalence of dynamic quantum circuits. 
Note that the latter
%equivalence checking algorithm in~\cite{hong2020tensor} 
cannot deal with dynamic quantum circuits. So we 
%extract 
isolate from our algorithms the 
%part of 
module for constructing TDD representations of dynamic circuits, and compare it with the 
%part
module
of the algorithm in~\cite{hong2020tensor} which constructs TDD representations of the corresponding conventional circuits. This comparison is fair, as both our algorithms and the algorithm in~\cite{hong2020tensor} need to construct two such TDDs and, once the TDDs are computed, the time/memory consumption of equivalence checking is negligible.

% Since the algorithms essentially just generate two TDDs for two equivalent conventional circuits, we only compare our algorithms with the procedure for calculating the separate TDD of the corresponding conventional circuit.

\begin{table}[]
\caption{Experiment results}
\centering
\scalebox{0.7}{
\begin{tabular}{|c|l|l|l|l|l|l|l|l|}
\hline
\multicolumn{1}{|l|}{\multirow{2}{*}{}}                           & \multicolumn{1}{c|}{\multirow{2}{*}{Benchmarks}} & \multicolumn{3}{c|}{Basic Alg.}                                         & \multicolumn{2}{c|}{Optimised Alg.}    & \multicolumn{2}{c|}{Alg. in~\cite{hong2020tensor}} \\ \cline{3-9} 
\multicolumn{1}{|l|}{}                                            & \multicolumn{1}{c|}{}                            & \multicolumn{1}{c|}{tdd\_time(s)} & \multicolumn{1}{c|}{time(s)} & m\_nodes & \multicolumn{1}{c|}{time(s)} & m\_nodes & \multicolumn{1}{c|}{tdd\_time(s)}   & nodes   \\ \hline
\multirow{21}{*}{\begin{tabular}[c]{@{}c@{}}m-\\ eq\end{tabular}} & qft\_2                                           & 0.00                           & 0.01                      & 8          & 0.01                      & 6          & 0.00                        & 7       \\ \cline{2-9} 
                                                                  & qft\_3                                           & 0.01                           & 0.03                      & 15         & 0.01                      & 10         & 0.01                        & 15      \\ \cline{2-9} 
                                                                  & qft\_4                                           & 0.02                           & 0.04                      & 31         & 0.02                      & 18         & 0.02                        & 31      \\ \cline{2-9} 
                                                                  & qft\_5                                           & 0.05                           & 0.08                      & 63         & 0.03                      & 34         & 0.03                        & 63      \\ \cline{2-9} 
                                                                  & qft\_6                                           & 0.08                           & 0.13                      & 127        & 0.06                      & 66         & 0.05                        & 127     \\ \cline{2-9} 
                                                                  & qft\_7                                           & 0.16                           & 0.27                      & 255        & 0.10                      & 130        & 0.09                        & 255     \\ \cline{2-9} 
                                                                  & qft\_8                                           & 0.32                           & 0.52                      & 511        & 0.18                      & 258        & 0.19                        & 511     \\ \cline{2-9} 
                                                                  & qft\_9                                           & 0.51                           & 0.87                      & 1023       & 0.34                      & 514        & 0.34                        & 1023    \\ \cline{2-9} 
                                                                  & qft\_10                                          & 1.45                           & 2.65                      & 2047       & 0.72                      & 1026       & 1.19                        & 2047    \\ \cline{2-9} 
                                                                  & qft\_11                                          & 2.79                           & 4.59                      & 4095       & 1.34                      & 2050       & 1.79                        & 4095    \\ \cline{2-9} 
                                                                  & qft\_12                                          & 7.71                           & 11.21                     & 8191       & 2.96                      & 4098       & 3.46                        & 8191    \\ \cline{2-9} 
                                                                  & qft\_13                                          & 13.74                          & 19.83                     & 16383      & 5.81                      & 8194       & 6.03                        & 16383   \\ \cline{2-9} 
                                                                  & qft\_14                                          & 28.01                          & 39.80                     & 32767      & 10.59                     & 16386      & 11.66                       & 32767   \\ \cline{2-9} 
                                                                  & qft\_15                                          & 53.76                          & 75.08                     & 65535      & 20.87                     & 32770      & 21.18                       & 65535   \\ \cline{2-9} 
                                                                  & qft\_16                                          & 119.31                         & 158.57                    & 131071     & 42.54                     & 65538      & 39.00                       & 131071  \\ \cline{2-9} 
                                                                  & PE\_2                                            & 0.01                           & 0.03                      & 20         & 0.01                      & 11         & 0.01                        & 20      \\ \cline{2-9} 
                                                                  & PE\_3                                            & 0.03                           & 0.07                      & 58         & 0.03                      & 23         & 0.03                        & 58      \\ \cline{2-9} 
                                                                  & PE\_4                                            & 0.16                           & 0.28                      & 180        & 0.06                      & 47         & 0.09                        & 180     \\ \cline{2-9} 
                                                                  & PE\_5                                            & 0.31                           & 0.84                      & 614        & 0.11                      & 95         & 0.4                         & 614     \\ \cline{2-9} 
                                                                  & PE\_6                                            & 1.32                           & 2.41                      & 2248       & 0.19                      & 191        & 0.56                        & 2248    \\ \cline{2-9} 
                                                                  & PE\_7                                            & 1.01                           & 3.34                      & 2664       & 0.32                      & 383        & 1.80                        & 762     \\ \hline
\multirow{5}{*}{\begin{tabular}[c]{@{}c@{}}q-\\ eq\end{tabular}}  & Bitflip                                          & 0.01                           & 0.06                      & 108        & 0.03                      & 50         & 0.03                        & 108     \\ \cline{2-9} 
                                                                  & Phaseflip                                        & 0.01                           & 0.10                      & 108        & 0.06                      & 51         & 0.06                        & 108     \\ \cline{2-9} 
                                                                  & Teleportation                                    & 0.00                           & 0.03                      & 20         & 0.02                      & 14         & 0.01                        & 20      \\ \cline{2-9} 
                                                                  & State\_inject\_S                                    & 0.00                           & 0.01                      & 10         & 0.01                      & 8          & 0.01                        & 10      \\ \cline{2-9} 
                                                                  & State\_inject\_T                                   & 0.01                           & 0.02                      & 10         & 0.01                      & 8          & 0.01                        & 10      \\ \hline
\end{tabular}}
\begin{tablenotes}
\item[1]* The `nodes' column records the numbers of nodes in the TDDs of the  circuit and the `m\_nodes' columns record the maximum numbers of nodes of the TDDs constructed in the calculation process.
\item[2]* The `tdd\_time' columns record the time for constructing the TDDs for the circuits, and the  `time' columns record the time for running the algorithm (including TDD construction time).
\end{tablenotes}
\label{experiment-results}
\end{table}

{\bf Results.}
Table \ref{experiment-results} gives the experiment results, where columns `Basic Alg.' and `Alg. in~\cite{hong2020tensor}' record the time and memory consumption in generating TDD representations for a dynamic circuit by our basic algorithms and the corresponding conventional circuit by the compared algorithm, respectively. From the table we can see that our basic algorithms run as effectively as the equivalence checking algorithm for conventional quantum circuits,
%The time consumption of our algorithms are comparable (mostly within 3x) with that for generating the TDD of the corresponding conventional quantum circuit, 
and the maximum number of the nodes of the TDDs constructed in the calculation process is very close to the number of nodes of the final TDD representation of the circuit. 
% This means that our \lsj{optimised?} algorithms are almost optimal in terms of the memory space in such a scheme. \lsj{This is confusing. What about the optimised algorithms}
%since the final TDD representation has to be calculated unless you use other schemes. 
%Therefore, the performance of our algorithms %will not be worse than 
%are comparable with that for conventional circuits.
%equivalence checking algorithm.

\subsection{Comparison of the Optimised and Basic Algorithms}

%{\bf Baseline.} 
We also implemented two optimised algorithms for m- and q-equivalence checking, that is, the qubit-by-qubit equivalence checking method proposed in Sec.~\ref{subsec:opt}, and then compare them with the basic algorithms.
%, which determines the equivalence by directly calculating the TDDs of the two dynamic quantum circuits and comparing them.

{\bf Results.} Results of the optimised methods are recorded in column `Optimised Alg.' of Table~\ref{experiment-results}, where we can see that the optimised algorithms can significantly reduce the time/space consumption of the equivalence checking process. Consider the circuit qft\_16 as an example. The basic algorithm takes 119 seconds with the maximal number of nodes appeared during the process being 131,071. In comparison, the optimised algorithm only takes 42.54 seconds with a maximum of 65,538 nodes. Thus, this optimisation method can really help in reducing the time and space consumption of the algorithms.

% In conclusion, our algorithm can accomplish the equivalence checking of dynamic quantum circuits and the effectiveness of our algorithm is the same as checking the equivalence of two conventional circuits. A part-by-part scheme is supposed to be able to help to improve the performance, and just a very simple qubit-by-qubit scheme can really work.

\section{Conclusion}\label{sec:conclusion}
Dynamic quantum circuits have been introduced as an effective method for executing quantum algorithms on near-term quantum devices. In this paper, we gave a formal definition of dynamic quantum circuits and proposed to characterise their functionalities in terms of ensembles of linear operators. Based on this novel semantics, two dynamic quantum circuits are equivalent if they have the same functionality. We further introduced two special kinds of equivalences for dynamic quantum circuits and implemented two equivalence checking algorithms based on the tensor decision diagram representations of dynamic quantum circuits. Experiments show that these algorithms are as efficient as the corresponding algorithms for conventional quantum circuits. 

In the future, we will explore methods like employing an optimal contraction order to further improve the performance of our equivalence checking algorithms. Another open question is to devise effective algorithms for checking the equivalence of general dynamic quantum circuits, which requires calculating partial traces and thus advanced tensor network contraction techniques like that introduced in \cite{Pan+approximate_tensor_contraction} may help.

%The equivalence of dynamic quantum circuits is explored and two kinds of equivalence are defined depending on whether we are concerned with the output quantum states or the final measurement outcomes. We give two algorithms for checking the equivalence of dynamic quantum circuits by transform the circuits to tensor networks. Our algorithm is shown to be as efficient as the usual decision diagram-based equivalence checking method for conventional quantum circuits. In the future, we will explore if other optimisation methods that have been used in the equivalence checking of conventional quantum circuits can be applied to check the equivalence of dynamic quantum circuits and improve its performance further. We will also explore if there are more efficient algorithms for checking the equivalence of general dynamic quantum circuits.

% \lsj{Algorithms for general dQC equivalence.} 

\bibliographystyle{IEEEtran}
\bibliography{references}

\end{document}